\newcommand{\mAJ}[3]{ #1, AJ, {#2}, #3}
\newcommand{\mApJ}[3]{ #1, ApJ, {#2}, #3}

\newcommand{\mApJS}[3]{ #1, ApJS, {#2}, #3}
\newcommand{\mAeA}[3]{ #1, A\&A, {#2}, #3}

\newcommand{\mAeAS}[3]{ #1, A\&AS, {#2}, #3}

\newcommand{\mARAeA}[3]{ #1, ARAA, {#2}, #3}

\newcommand{\mMN}[3]{ #1, MNRAS, {#2}, #3}

\def\BGA{\begin{array}}
\def\EDA{\end{array}}
\def\BGD{\begin{description}}
\def\EDD{\end{description}}
\def\BGF{\begin{figure}}
\def\EDF{\end{figure}}
\def\BGE{\begin{equation}}
\def\EDE{\end{equation}}
\def\BGC{\begin{center}}
\def\EDC{\end{center}}
\def\BGT{\begin{tabular}}
\def\EDT{\end{tabular}}
\def\BGIT{\begin{itemize}}
\def\EDIT{\end{itemize}}
\def\BGEN{\begin{enumerate}}
\def\EDEN{\end{enumerate}}
\def\and{{\&}}

\def\dex1{\mbox{dex}}
\def\dex{\hbox{\rm dex}}

\def\eg1{{e.g.\/}}

\def\gp{\hbox{\rlap{\hbox{.}}\raise 5.truept \hbox{{\small$\circ$}}}}
\def\gradip{\hbox{\rlap{\hbox{.}}\raise 5.truept \hbox{{\small $\circ$}}}}
\def\orep{\hbox{\rlap{\hbox{.}}\raise 5.truept \hbox{{\small $h$}}}}

\def\magcir{\ \raise-2.truept\hbox{\rlap{\hbox{$\sim$}}\raise5.truept
\hbox{$>$}\ }}

\def\mincir{\ \raise-2.truept\hbox{\rlap{\hbox{$\sim$}}\raise5.truept
\hbox{$<$}\ }}

\def\underline{}

\def\lsim{\,\lower2truept\hbox{${< \atop\hbox{\raise4truept\hbox{$\sim$}}}$}\,}
\def\gsim{\,\lower2truept\hbox{${> \atop\hbox{\raise4truept\hbox{$\sim$}}}$}\,}

\def\s{\,{\rm s}}

\def\logre{\mbox{$\log(r_{\rm e})$}}
\def\s_fit{\mbox{$\sigma_{\rm bf}$}}

\def\dex{\mbox{dex}}

\def\eg{{\it e.g}}

\def\$\sigma$c{\mbox{$\$\sigma$_c$}}

\def\gp{\hbox{\rlap{\hbox{.}}\raise 5.truept \hbox{{\small$\circ$}}}}

\def\lsim{\, \lower2truept\hbox{${< \atop\hbox{\raise4truept\hbox{$\sim$}}}$}\,}
\def\gsim{\, \lower2truept\hbox{${> \atop\hbox{\raise4truept\hbox{$\sim$}}}$}\,}
\documentstyle[draft]{mn}
\title[CMB Anisotropies from Extragalactic Sources] 
{Extragalactic Source Counts and Contributions
to the Anisotropies of the Cosmic Microwave Background.
Predictions for the Planck Surveyor mission. }

\author[L. Toffolatti et al.]
{L. Toffolatti $^{1,2}$, F. Arg\"ueso G\'omez $^3$, G. De Zotti $^1$, P. 
Mazzei $^1$,\and
A. Franceschini $^4$, L. Danese $^5$ and C. Burigana $^6$
\\
$^{1}$Osservatorio Astronomico, Vicolo dell'Osservatorio 5, I-35122 Padova, 
Italy \\
$^{2}$Departamento de F\'\i{sica}, Universidad de Oviedo, c.le Calvo Sotelo 
s/n, E-33007 Oviedo, Spain \\
$^{3}$Departamento de Matem\'aticas, Universidad de Oviedo, c.le Calvo Sotelo 
s/n, E-33007 Oviedo, Spain \\
$^{4}$Dipartimento di Astronomia, Universit\`a di Padova, Vicolo 
dell'Osservatorio 5, I-35122 Padova, Italy \\
$^{5}$SISSA - International School for Advanced Studies, Via Beirut 2--4, 
I-34013 Trieste, Italy \\
$^{6}$Istituto TESRE, Consiglio Nazionale delle Ricerche, Via Gobetti 101, 
I-40129 Bologna, Italy \\}

\date{Accepted 1998 January 15. Received 1997 December 19; 
in original form 1997 September 29} 

\begin{document}
\label{firstpage}

\maketitle

\begin{abstract}

We present predictions for the counts of extragalactic sources, 
the contributions to 
fluctuations and their angular power spectrum in each channel foreseen for
the Planck Surveyor (formerly COBRAS/SAMBA) mission. The contribution 
to fluctuations due to 
clustering of both radio and far--IR sources is found to be generally
small in comparison with the Poisson term;
however the relative importance of
the clustering contribution increases and may eventually 
become dominant if sources are 
identified and subtracted down to faint flux limits. 
The central Planck frequency bands are expected to be ``clean'':
at high galactic latitude ($\vert b \vert >20^\circ$), where the
reduced galactic noise does not prevent the detection of the
extragalactic signal, only a tiny fraction of pixels is found to be
contaminated by discrete extragalactic sources.
Moreover, the ``flat'' angular power spectrum of fluctuations due to
extragalactic sources substantially differs from that of primordial
fluctuations; therefore, the removal of contaminating signals is
eased even at frequencies where point sources give a sizeable contribution
to the foreground noise.
%
%
\end{abstract}

\begin{keywords}
cosmic microwave background: anisotropies,
foregrounds -- radio and far--IR sources: counts, spatial distribution
-- galaxies: evolution.
\end{keywords}

\section{Introduction}
%


The long-sought COBE/DMR discovery (Smoot et al. 1992) of anisotropies of the 
Cosmic Microwave Background (CMB) has stimulated an outburst of activity in 
the field. Many new experiments have been undertaken and detections of CMB 
fluctuations have been reported on several angular scales 
(see White, Scott \& Silk 1994; Smoot 1997 for reviews). 

An unavoidable fundamental limitation to these measurements is set by 
astrophysical foregrounds.
Due to the large beam size, the COBE/DMR data are, to some extent, 
contaminated by the galactic emission (see, e.g., Kogut et al. 1996a,b)
whereas they are basically unaffected by extragalactic
foreground sources (Banday et al. 1996; Kogut et al. 1994).
On the other hand, extragalactic foregrounds are a major problem for 
high resolution experiments reaching the sensitivity of 
$\Delta T/T \simeq 10^{-6}$, as the recently selected ESA's
COBRAS/SAMBA -- now Planck Surveyor -- and NASA's MAP satellite missions, 
as well as some balloon--borne experiments, do. 

This paper presents 
a thorough analysis of the extragalactic foreground contribution to 
small scale fluctuations, over the full wavelength range
from $\sim$ 1 cm down to $\sim$300 $\mu$m covered by the Planck mission. 
The ten channels currently foreseen for the experiment as well as some 
relevant details on the payload characteristics are given in Table 1.

Estimates of temperature fluctuations due to a Poisson distribution of 
extragalactic sources have been worked out by 
Franceschini et al. (1989, 1991), Wang (1991), Blain \&
Longair (1993), Toffolatti et al. (1995), Danese et al. (1996), 
Tegmark \& Efstathiou (1996), and Gawiser \& Smoot (1997).

In the frequency range of interest here, there are important contributions 
both from the high frequency tail of the spectrum of radio sources and 
from the long wavelength portion of the dust emission in galaxies. 
The present work improves on previous analyses for both populations.

We adopt updated models for the evolution of galaxies 
which account for the faint number 
counts at $60\,\mu$m, recently reassessed by Bertin, Dennefeld \& Moshir 
(1997) as well as for the preliminary estimates of deep counts at  $6.7$ 
and $15\,\mu$m (Oliver et al. 1997) obtained from long exposures with the 
camera of the Infrared Space Observatory (ISO) and at $170\,\mu$m with
the ISO long-wavelength photometer (Kawara et al. 1997),
as discussed by Franceschini et al. (1997).
Our models deal in a self-consistent way with the evolution
of the spectral energy distribution of galaxies from UV to far-IR 
wavelengths (Mazzei, Xu \& De Zotti 1992;  Mazzei, De Zotti \& Xu 1994), 
so that we can take advantage of surveys in a broad wavelength range to 
test them. Also, they naturally predict a relationship 
between the dust temperature 
distribution and the intensity of star-formation activity, in 
the sense that more actively star-forming galaxies have warmer dust 
emission spectra, consistent with the observed correlation between 
the ratio $f_{60\mu{\rm m}}/f_{100\mu{\rm m}}$ and the bolometric 
luminosity of IRAS selected galaxies (cf. e.g. Sanders \& Mirabel 1996). 

Moreover, we have explored various possibilities for the shape of the 
high frequency spectra of radio selected sources to take into account 
that the situation is likely 
rather complex: on one side, the non thermal emission spectrum of compact 
sources is 
expected to steepen or even to break at mm wavelengths; on the other hand, 
substantial contributions at these wavelengths probably due to dust 
emission have been reported for several objects. 

We also give a quantitative estimate of the additional contributions to 
fluctuations due to clustering. 

The outline of the paper is as follows. In Section 2 we briefly review the
basic formalism. In Section 3 we deal with source counts and their 
extrapolation to the Planck bands. Our results are presented 
and discussed in Section 4. Finally, in Section 5, we summarize our 
main conclusions.

Throughout this paper we will adopt an Einstein-de Sitter ($\Omega =1$) 
cosmology with 
a Hubble parameter of $H_0=50\,\hbox{km}\,\hbox{s}^{-1}\,\hbox{Mpc}^{-1}$. 

\begin{table*}
\centering
\caption{\bf Planck Surveyor -- Model Payload Characteristics}
\normalsize

\bigskip

\begin{tabular}{|l||cccc|cccccc|} 
\hline
Telescope & \multicolumn{10}{c|}{1.5 m Diam. Gregorian; system emissivity 1\% }\\
     & \multicolumn{10}{c|}{Viewing direction offset 90$^\circ$ from spin axis}\\
\hline
Instrument        & \multicolumn{4}{|c|}{LFI} & \multicolumn{6}{|c|}{HFI} \\
\hline
Detector Technology & \multicolumn{4}{|c|}{HEMT receiver arrays} & 
\multicolumn{6}{|c|}{Bolometer arrays} \\
\hline
Detector Temperature       & \multicolumn{4}{c|}{$\sim$ 20 K }& 
\multicolumn{6}{c|}{0.1-0.15 K} \\
\hline
Center Frequency (GHz) & 30  & 44 & 70 & 100 & 100 & 143 & 217 & 353 & 545 & 857 \\
\hline
Wavelength (mm) & 10 & 6.8 & 4.3 & 3.0 & 3.0 & 2.1 & 1.38 & 0.85 & 0.55 & 0.35 \\
\hline
Bandwidth (GHz) & 6.0 & 8.8 & 13 & 20 & 37 & 53 & 80 & 131 & 202 & 317 \\
\hline
Number of Detectors & 4 & 6 & 12 & 34 & 4 & 12  &  12 &  6  &  6 & 6 \\
\hline
Angular Resolution & 33 & 23 & 14 & 10 & 10.6 & 7.4 & 4.9 & 4.5 & 4.5 & 4.5 \\
 ~~~(FWHM, arcmin) &      &  &   &  &  &  & &  &   & \\
\hline
Optical Transmission        & 1  & 1  & 1  & 1  & 0.3 & 0.3 & 0.3 & 0.3 & 0.3 & 0.3\\
\hline
$\langle {\Delta T/ T}\rangle$ Sensitivity per res. element & 1.6 & 2.4 & 3.6 & 
4.3 & 1.81 & 2.1 & 4.6 & 15.0 & 144. & 4630 \\
~~~(1 year, 1$\sigma$, $10^{-6}$ units) & & & & & & & & & &\\
\hline
\end{tabular}
\end{table*}

\section{Basic Formalism}

All the estimates presented in this paper are based on the assumption
of ``point-like'' sources.
As shown by Rowan--Robinson \& Fabian (1974), this is a good approximation
as far as the angular sizes of sources do not exceed the beam width.
This is generally the case for the Planck mission 
(FWHM$\geq 10'$ at $\nu\leq 140$ GHz and
FWHM$\geq 4'.4$ for the high frequency channels; see Table 1); the pixels 
contaminated by the few bright very extended sources will be anyway removed. 

\subsection{Sky fluctuations from randomly distributed sources}

The problem has been extensively discussed in the literature 
(Scheuer 1957, 1974; Condon 1974; Franceschini et al. 1989); 
we recall here only the basic points. 
A useful estimate of the fluctuation level generated by
randomly distributed sources is provided by the second moment, $\sigma$, 
of the distribution of the mean number, $R(x)$, of responses 
$x= S f(\psi)$ to sources of flux $S$ located at an angular distance
$\theta$ from the beam axis, 
$f(\psi)$ being the angular power pattern of the detector, for which we 
adopt a gaussian shape with FWHM $\theta_0$:
\begin{equation}
\sigma^2 = \int_0^{x_c}x^2R(x)dx = \pi\theta_0^2I(x_c) , 
\end{equation}
with
\begin{equation}
I(x_c)  = \int_0^{x_c}dx x^2 \int_0^{\infty}d\psi
N\left({x\over f(\psi )}\right) \exp(4\psi \ln 2), 
\end{equation}
where $\psi \equiv (\theta/\theta_0)^2$, 
$N(S)$ are the differential source counts per steradian at 
a given frequency $\nu$
and $x_c$ is the limit above which 
a source is considered to be individually detected.
Following Condon (1974), we set $x_c=q \sigma$; $q$ is usually taken in 
the range 3--5. We adopt $x_c=5\sigma$ throughout this paper. 
The confusion standard deviation $\sigma$ is related to the rms brightness
temperature fluctuations ($\Delta T/T_{\rm rms}\equiv \langle 
(\Delta T/T)^2\rangle^{1/2}$), at the wavelength $\lambda$ by: 
\begin{eqnarray}
\left({\Delta T\over T}\right)_{\rm rms} & = &
{\lambda^2\sigma\over 2kT\omega_{\rm eff,I}} \left[\exp\left({h\nu\over
kT}\right)-1\right]^2 \nonumber \\
& \times &\exp\left(-{h\nu\over kT}\right)\displaystyle/
\left({h\nu\over kT}\right)^2 , 
\end{eqnarray}
where 
\begin{equation}
\omega_{\rm eff,I}=\int d\omega f(\psi) 
\end{equation}
is the effective beam area, $d\omega$ being the 
solid angle differential element (De Zotti et al., 1996a).
In terms of the angular distance $\theta$ from the beam axis equation (4)
reads $\omega_{\rm eff,I}=\pi\int d\theta^2 f(\theta)$
(Franceschini et al., 1989).
%
$T =2.726\ K$ (Mather et al. 1994) is the brightness temperature of the CMB.
In terms of intensity fluctuations we have:
\begin{equation}
\left({\Delta I_{\nu}\over I_{\nu}}\right)_{\rm rms} = 
{\sigma \over I_{\nu}\omega_{\rm eff,I}} . 
\end{equation}
In the Rayleigh-Jeans region $\Delta I_{\nu}/I_{\nu} \simeq \Delta T/T$
while at the peak of the intensity of the CMB, ($\lambda\simeq 1$ mm),
$\Delta I_{\nu}/I_{\nu} \simeq 3 \Delta T/T$.

\subsection{Fluctuations due to clustered sources}

Clustering decreases the effective number of objects in randomly
distributed cells and, consequently, enhances the cell-to-cell fluctuations
(Peebles 1980; Barcons \& Fabian 1988). The analysis of a complete
sample of nearby ($z<0.1$) radiogalaxies selected at 1.4 GHz
(Peacock \& Nicholson 1991) has shown that, at least in a particular
range of radio power, sources are strongly clustered 
(correlation length $r_0\simeq 22 (H_0/50)^{-1}\,\hbox{Mpc}$). 
Evidences of a strong angular correlation were found by Kooiman, Burns 
\& Klypin (1995) in the Green Bank 4.85 GHz catalog.  
More recently, Loan, Wall \& Lahav (1997) estimated the 
angular two--point correlation function of sources selected
at 4.85 GHz, by combining the Green Bank and Parkes--MIT--NRAO surveys. 
For an evolution index of the correlation function (Peebles 1980) 
in the range $-1.2 < \epsilon < 0$ they found $26\,\hbox{Mpc} < (H_0/50) r_0 < 
36\,\hbox{Mpc}$. 

Significant clustering was also established for far--IR selected sources, 
on a scale 
$r_0\sim 8(50/H_0)$ Mpc (Saunders, Rowan--Robinson \& Lawrence, 1992) 
slightly smaller than for optically selected galaxies. 

The contribution of clustering to intensity fluctuations is straightforwardly
obtained from the angular correlation function $C(\theta_\star)$, as a
function of the angular separation $\theta_\star$, setting
$\theta_\star =0$ (see, e.g., De Zotti et al. 1996a). If the clustering scale 
is much smaller than the Hubble radius, we have:
\begin{equation}
\left(\Delta I \over I \right)_{cl} = \Gamma(\theta_\star\rightarrow 0)= \sqrt
{ C(\theta_\star\rightarrow 0)\over {\langle I\rangle }^2}
\end{equation}
where ${\langle I\rangle}$ is the mean background intensity and
\begin{eqnarray}
C(\theta_\star )&= & {\left(c \over 4 \pi H_o\right)^2}
\int d\omega f(\vartheta , \varphi ) \int d\overline \omega  
f(\overline \vartheta  , \overline \varphi ) \nonumber \\
&\times & \int_{z_m(L_{min},S_l)}^{z_{max}} dz  {j_{\rm eff}^2(z) 
\over (1 + z)^6 (1 + \Omega z)} \nonumber \\
&\times & \int_{\max \left[z_m - z, -\Delta 
(r_{\rm max})\right] }^{\min \left[z_{max} - z, \Delta(r_{\rm max})\right] }
d(\delta z) \xi (r,z). 
\end{eqnarray}
Here $\xi (r,z)= h(z)\xi_0 (r)$ is the two--point spatial correlation
function, with $\xi_0 (r)=(r_0/r)^{1.8}$. For the clustering evolution 
function we adopt the usual simple expression $h(z)=(1+z)^{-(3+\epsilon )}$. 
$\Delta (r_{max})$
is the value of $\delta z$ corresponding to the
maximum scale of clustering, $S_l$ is the adopted flux limit and 
\begin{equation}
j_{eff}(z) = \int_{L_{\rm min}}^{\min\left[ L_{\rm max}, L(S_l,z) \right] } 
d\log L \ L \ n_c(L,z) K(L,z) 
\end{equation}
is the effective volume emissivity, $n_c(L,z)$ being the comoving number
density of sources and $K(L,z)$ the K-correction factor. 
The pairs of angles $(\vartheta , \varphi )$ 
and $(\overline \vartheta  , \overline \varphi )$ define two directions 
separated by an angle $\theta_\star$.
It is worth stressing that, while for differential source counts 
[$N(S) \propto S^{-\beta}$] with 
slope $\beta < 3$ the Poisson noise is dominated by the
sources just below the detection threshold, $S_{l}$, the main
contribution due to clustering always comes from the faintest sources
which do actually cluster on the given angular scale. 
Therefore, lowering $S_{l}$ results in an increase of the relative 
importance of the clustering term in comparison with the Poisson one.

\subsection{Angular power spectrum analysis of the intensity fluctuations}

We consider, as usual, the spherical harmonic expansion of the sky temperature
fluctuations:
\begin{equation}
{\delta T\over T}(\vartheta ,\varphi) =\sum_{\ell=0}^\infty
\sum_{m=-\ell}^{\ell}{a_{\ell}^{m}}Y_{\ell}^m(\vartheta ,\varphi). 
\end{equation}
If the fluctuations are a stationary process, the angular power spectrum is 
independent of $m$ (see Peebles 1993, p. 517), so that 
\begin{equation}
C_\ell(\nu) = {1\over 2\ell +1}  \sum_{m=-\ell}^{\ell}
\langle\vert{a_{\ell}^{m}}\vert^2\rangle =  
\langle\vert{a_{\ell}^{0}}\vert^2\rangle. 
\end{equation}
with 
\begin{equation}
a_{\ell}^0=\int{{\delta T\over T}(\vartheta ,\varphi)Y_l^0(\vartheta ,
\varphi)d\Omega}, 
\end{equation}
or, in terms of Legendre polynomials,
\begin{equation}
a_{\ell}^0=\int_0^{2\pi}\int_0^{\pi}{\delta T\over T}(\vartheta
,\varphi)\sqrt{{2\ell +1\over
4\pi}}P_{\ell}(\cos\vartheta)\, \sin\vartheta\, d\vartheta d\varphi.
\end{equation}
The estimate of $C_{\ell}(\nu)$ is obtained  by dividing the sky
in [$N\times N$] ``equal area'' cells
(the cell dimension being $1.5\times 1.5$ arcmin$^2$)
$\left[\vartheta_{i-1},\vartheta_i\right]\times\left[\varphi_{j-1},\varphi_j
\right]$
so that in every cell the fluctuation $\delta T/T$ can be considered
constant. Then
\begin{equation}
a_{\ell}^0=\sum_{i=1}^N\sum_{j=1}^N\left({\delta T\over T}\right)_{i,j}
\sqrt{{2\ell+1\over 4\pi}}\,{2\pi \over N}\,{A_{\ell(i)}\over{2\ell+1}}, 
\end{equation}
where
\begin{eqnarray}
\lefteqn{{A_{\ell(i)}\over{2\ell+1}}  = 
\int_{\vartheta_{i-1}}^{\vartheta_i} P_{\ell}(\cos\vartheta)\,
\sin\vartheta\, d\vartheta  =  {1\over{2\ell+1}} [P_{\ell-1}
\left(\cos\vartheta_i\right) - } \nonumber \\
 & & P_{\ell-1}\left(\cos\vartheta_{i-1}\right) 
-P_{\ell+1}\left(\cos\vartheta_i\right)+P_{\ell+1}\left(\cos\vartheta_{i-1}
\right)]. 
\end{eqnarray}
The temperature fluctuations $\left(\delta T/T\right)_{i,j}$ in every cell
were calculated from simulations of the all--sky distribution of extragalactic 
point--like sources, based on 
our estimates of the source counts for each frequency channel of the 
Planck mission, rather than from the formulae derived by Tegmark \& Efstathiou 
(1996). This approach allows to calculate the angular power spectrum under 
more general conditions (in particular allowing for the effect of 
clustering) since it only requires that
{\it the overall source distribution be statistically isotropic.}
We have checked that, assuming a Poisson distribution and using the source 
counts estimated by Tegmark \& Efstathiou (1996), our method reproduces 
their results.  


Following Tegmark \& Efstathiou (1996) we then computed the quantities
\begin{equation}
\delta T_{\ell}(\nu)=[\ell(\ell + 1)C_{\ell}(\nu)/2\pi]^{1/2}. \label{eq:dT}
\end{equation}
%

\section{Source counts and their extrapolation
to millimeter and sub--mm wavelengths}

\subsection{Counts of radio sources}

Deep VLA surveys have allowed to extend direct determinations of radio 
source counts down to $\mu$Jy levels at 1.41, 4.86 and 8.44 GHz. At these 
frequencies counts now cover about 7 orders of magnitude in flux and 
reach an areal density of several sources arcmin$^{-2}$. Therefore, 
at these frequencies, fluctuations can be determined directly from the counts 
down to angular scales much smaller than those of interest here. 

As shown by Figure 1, the model by Danese et al. (1987) provides a good 
fit to the available data, at least for $S > 100\,\mu$Jy. Particularly 
encouraging is the 
good agreement with the deep counts at 8.44 GHz (Windhorst et al. 1993; 
Partridge et al. 1997), which were produced several years after the model, 
indicating that the adopted distribution of spectral indices 
of sources was appropriate. 

Counts below $100\,\mu$Jy affect fluctuations due to radio sources on 
scales much smaller than those reachable by the Planck mission. 

The assumptions about source spectra are 
obviously the most critical ingredient for the purposed extrapolation of 
the observed counts to much higher frequencies.

Radio loud AGNs, including ``flat''-spectrum radiogalaxies, quasars, 
BL-Lacs, mostly at substantial $z$, are expected to dominate the counts in 
the Planck low frequency channels for $S \gsim 1\,$mJy and, correspondingly, 
the contribution of extragalactic sources to fluctuations on the angular 
scales of interest here.

The analysis of Impey \& Neugebauer (1988), giving the spectral indices of 
162 blazars, and the quasi-simultaneous observations of 176 bright compact
sources done by Edelson (1987), have shown that compact sources
have ``flat'' spectra ($S(\nu)\propto \nu^{-\alpha}$, with $\alpha \simeq 0$,
although with some scatter) at least up to $\sim 100$ GHz. This conclusion 
agrees with previous analyses (Owen, Spangler \& Cotton 1980), which showed 
that the spectral indices of most strong flat--spectrum radio sources 
keep flat ($\alpha < 0.3$) in the range 1 to 100 GHz. 

At still higher frequencies, a steepening or even a spectral break of the 
synchrotron emission is expected. On the other hand, the presence of 
excess mm emission over extrapolations from cm wavelengths has been 
established by Knapp \& Patten (1991) for a sample of nearby radio galaxies. 
Observations of large mm fluxes attributed to dust emissions have been 
reported for several distant radio galaxies (see Mazzei \& De Zotti, 1996 
and references therein). The inferred dust masses are 1--2 orders of magnitude 
higher than found for nearby radio galaxies. The two components (synchrotron 
and dust emission) may well have different evolution properties.

In view of the uncertainties on the spectra mentioned above, our estimates
of counts and fluctuations due to radio selected sources have been calculated
as it follows. We adopt the Danese et al. (1987) simple luminosity evolution
model and three different choices for the average spectral index of
``flat--spectrum'' compact sources:
$a)$ $\alpha =-0.3$, $b)$ $\alpha=0.0$, and $c)$ $\alpha=0.3$ at
$20\lsim \nu\lsim 200$ GHz, with a steepening to $\alpha =0.7$ at higher
frequencies; below 20 GHz we have set $\alpha =0$. 
As for ``steep''--spectrum sources (elliptical, S0 and starburst
galaxies), whose contribution to source counts is actually minor in the
whole frequency range of interest here, the radio power -- spectral index
relation determined by Peacock and Gull (1981) has been adopted.
The maximum and minimum number of expected radio sources in each Planck 
channel, given in Table 2, correspond to the two extreme values of $\alpha$ 
for compact sources.

Holdaway et al. (1994) carried out sensitive 90 GHz observations 
of a sample selected at 5 GHz and observed at 8.4 GHz. They derived a 
distribution of spectral indices between 8.4 GHz and 90 GHz which allowed 
them to extrapolate to 90 GHz the 5 GHz source counts. They estimate 
that, over the entire sky, there are 178 sources with $S_{\rm 90\,GHz} > 
1\,$Jy, almost a factor of 2 less than 
predicted by our model. However, as discussed by these authors, 
their estimate is somewhat below the number of known sources 
with $S_{\rm 90\,GHz} > 1\,$Jy, so that it should be viewed as a lower 
limit. 

On the other hand, our estimate may miss a population of sources with 
strongly inverted radio spectra. However, the analysis by 
Condon et al. (1995) of a large sample of extragalactic sources detected 
by IRAS at $\lambda =60\,\mu$m
and identified with VLA radio source catalogues at 4.85 GHz, found no evidence 
of a significant population of sources with spectra rising steeply from 
centimeter to millimeter wavelengths.   

\subsection{Counts of far--IR sources }

Both in the case of normal and of many active galaxies (as far as 
we can tell, based on the very limited information currently available), 
at wavelengths shorter than a few mm (in the rest frame), dust emission 
rapidly overwhelms the radio emission. 
Due to the very steep increase with frequency of the dust 
emission spectrum at mm and sub-mm wavelengths ($\alpha \simeq -3.5$) 
the wavelength at which dust emission takes over does not change much 
between radio quiet and radio loud sources, in spite of the fact that 
the ratio of radio to far-IR emission for the former class 
is orders of magnitude lower than for the latter. 

As in the case of radio sources, estimates of the counts of far--IR sources 
in the frequency region covered by the Planck instruments are difficult 
because of the wide gap with the nearest wavelength ($60\,\mu$m) where 
the most extensive (yet relatively shallow) surveys exist. 
There is a considerable spread in the distribution of dust temperatures, 
so that the observed $1.3\,\hbox{mm}/60\,\mu$m flux ratios of galaxies span 
about a factor of 10 (Chini et al. 1995; Franceschini \& Andreani 1995; 
see also De Zotti et al. 1996b).
A tentative estimate of the luminosity function of galaxies at mm wavelengths 
based on a $1.25\,\hbox{mm}/60\,\mu$m bivariate luminosity distribution
has been presented by Franceschini, Andreani \& Danese (1997). 

Furthermore, the observational constraints on evolution properties of 
far-IR sources are very poor. They come essentially from IRAS $60\,\mu$m 
counts, which cover a limited range in flux and are rather uncertain 
at the faint end (Hacking \& Houck 1987; Gregorich et al. 1995; 
Ashby et al. 1996; Bertin, Dennefeld \& Moshir 1997).

From a theoretical point of view, there is a great deal of uncertainty on 
the physical processes governing galaxy formation and evolution. Some 
models assume that the comoving density of galaxies remained essentially 
constant after their formation, while they evolved in luminosity due to 
the ageing of stellar populations and the birth of new generations of stars 
(pure luminosity evolution; see, i.e., Franceschini et al., 1994 for
a most complete discussion on the subject). 
On the other hand, according to the hierarchical galaxy formation
paradigm, big galaxies are formed by coalescence of large numbers
of smaller objects (see, e.g. White 1996 and references therein).

Furthermore, evolution depends on an impressive number of unknown or
poorly known parameters: merging rate, star formation rate, initial 
mass function, galactic winds, infall, interactions, dust properties, 
etc.

Although the evolutionary history is highly uncertain, strong evolution
is expected in the far-IR/mm region particularly 
for early type galaxies since during their early phases they must 
have possessed a substantial metal enriched interstellar medium. 
This expectation is supported by evidences of large amounts of dust 
at high redshifts (cf. in particular the case of high-$z$ radiogalaxies: 
Mazzei \& De Zotti, 1996) 
and by the intensity of the isotropic sub--mm component reported by 
Puget et al. (1996) from an analysis of COBE/FIRAS data (see 
Burigana et al. 1997).

Moreover, the K-correction due to the steep increase with frequency 
of the dust emission spectrum below the peak, generally located  
at $\lambda \simeq 60$--$150\,\mu$m, strongly amplifies the 
evolutionary effects, that may thus be appreciable in the relatively 
shallow Planck surveys, at least in the highest frequency bands.

In view of the above, we adopt here two different descriptions of 
the cosmological evolutions of far--IR sources.

$i)$ Our reference model is an updated version of model C of 
Franceschini et al. (1994; {\it opaque model}), which provides a good fit 
(see Fig. 2) to the $60 \mu$m IRAS counts, as recently reassessed by Bertin et 
al. (1997) and to the 
far-IR extragalactic background spectrum estimated by Puget et al. (1996), 
as shown by Burigana et al. (1997). The model also predicts counts 
at $170\,\mu$m and at $850\,\mu$m which are in agreement, respectively, with 
the preliminary estimates by Kawara et al. (1997) using the ISO
long-wavelength photometer
and by Smail, Ivison \& Blain (1997) using the new bolometer array, SCUBA, 
on the JCMT (see Fig. 3).

Moreover, we find agreement with the recent estimate of the ``source''
confusion noise by Bertin et al. (1997; their Section 4.1), from the 
analysis of the IRAS Very Faint Source Sample (VFSS). Their estimated
value is $\sigma_{\rm conf}\sim 20$ mJy, about $2/3$ of the total
noise of $30.2\pm 1.2$ mJy (instrumental noise plus source confusion
added up in quadrature). By comparison, our reference model
gives a confusion standard deviation of $\sigma_{\rm conf} \sim 18-20$ mJy, 
with a gaussian beam pattern of FWHM$\sim 3$ arcmin and integrating
up to a source detection limit $x_c\simeq 100$ mJy.  

The updated {\it opaque model} adds a density evolution up to $z=2$
of late type galaxies
(spirals, irregulars, starburst) only, proportional $\exp(2\tau(z))$, with  
$\tau(z)=1-t(z)/t_0$, $t(z)$ being the age of the universe at redshift $z$ 
and $t_0$ its present age. The estimated $\Delta T/T_{\rm rms}$ levels
as well as the spatial power spectra of fluctuations
due to far-IR extragalactic sources
have all been computed exploiting this model (see Section 4).

On the other hand, the signal detected by Puget et al. (1996) may be
contaminated by cold dust emission from an extended Galactic halo (Fixsen et 
al. 1996). In that case our reference model would provide an upper limit 
to the evolution of extragalactic sources in the far-IR. 

$ii)$ A lower evolution rate in the far-IR is implied by the {\it moderate
extinction} model by Franceschini et al. (1994).
This second model predicts source counts which are still compatible
with the IRAS $60\,\mu$m counts and with the quoted estimated
counts at 170 $\mu$m by Kawara et al. (1997). On the other hand, they
slightly underestimate the number of sources quoted by Smail, Ivison \&
Blain (1997).

The estimated number counts, $N(>S_{\rm lim})_{\rm FIR}$,
of far-IR sources in the HFI frequency channels are a factor
$\simeq 1.2-1.8$ lower than the counts plotted in Fig. 4 (our reference
model) if $S_{\rm lim}\simeq 1$ Jy (the lower value, 1.2, applies to
the number counts ratio at 857 GHz whereas the value 1.8 is found for
the same ratio at 143 GHz). If we adopt the fainter
flux limit of $S_{\rm lim}\simeq 10^{-2}$ Jy then the model predicts 
counts $\simeq 1.4-2.2$ times lower than the updated model C.
The increasing difference between the estimated number counts,
if one moves shortward in frequency from 857 GHz down to 143 GHz,
is determined by the steep rise with increasing frequency of the
dust emission spectrum in the sub-mm domain which entails more 
important contributions to the counts from high redshift
galaxies.

The $\Delta T/T_{\rm rms}$ levels are
only $\sim$20-40\% lower than those predicted by the reference
model at the angular resolution limit of the HFI instrument.
At larger angular scales the difference is negligible since the
two models predict a very similar number of bright sources
(see before).
The maximum and minimum number of expected individually detected far-IR 
sources in the Planck surveys at $\vert b \vert > 20^\circ$ given in
Table 2 correspond to the predictions of the updated model C and of the
moderate extinction model.

The contribution of dust-rich galaxies to the radio counts (see Table
2) has been estimated assuming a linear relationship between the
far-IR flux at 60 $\mu$m (dominated by the starburst component)
and the radio centimetric flux. The ratio of the 60 $\mu$m to the
radio flux at 1.4 GHz was assumed to be $S_{60\,\mu m}/S_{1.4\,{\rm GHz}}
\simeq 140$ (see Helou et al., 1985).

The contributions of AGNs to the counts in the high frequency
channels (where, however, star-forming galaxies should be the 
dominant population) are even more uncertain, 
since they depend on the evolution of both the non-thermal component and 
the dust emission, which in turn depend on several unknown or 
poorly known factors, such as the evolution of the nuclear energy source, 
the effect of possible circumnuclear starbursts, the abundance, properties 
and distribution of dust, and so on. The IRAS survey data do not help much, 
since the detection rate of quasars was extremely low. 

Taking only into account the evolution of the 
non--thermal component, assumed to parallel that observed in X-rays, 
and adopting the spectral energy distributions used by Granato et al. (1997) 
for type 1 and type 2 AGNs, we 
expect a detection rate of radio quiet AGNs increasing with increasing 
frequency: only a few of them are expected at 1.4 mm, but their number 
should increase up to a few hundreds at the highest frequencies. In any case, 
the number of radio quiet AGNs is negligible compared with the 
number of far-IR galaxies.

\begin{table*}
\protect\small
\baselineskip=12pt
\noindent
\centering
\caption{\bf Instrumental and confusion noise estimates for the Planck mission 
and expected numbers of individually detectable sources at $\vert b \vert >
20^\circ$.}
\begin{tabular}{ccccccccccc}
\hline $\nu_{\rm eff}$ & $\lambda_{\rm eff}$ & beam & $\sigma_{\rm noise}$ & 
$\sigma_{\rm Gal}$ & $\sigma_{\rm conf}$ & $\sigma^{(1)}_{\rm CMB}$ 
& $S^{(2)}_{\rm lim}$ & $N(>S_{\rm lim})_{\rm radio}$ & 
$N(>S_{\rm lim})_{\rm FIR}$ \\ 
(GHz) & (mm) & (arcmin) & (mJy) & (mJy) & (mJy) & (mJy) & (mJy) & (8 sr) 
& (8 sr) \\ \hline
& & & & & & & & & \\
30 & 10.0 & 33 & 13 & 100 & 37 & 78 & 650 & 300-800 & 0-2 \\
44 & 6.8 & 23 & 19 & 45 & 17 & 80 & 480 & 450-1600 & 3-10 \\
70 & 4.3 & 14 & 25 & 15 & 8 & 68 & 330 & 600-2500 & 6-18 \\ 
100 & 3.0 & 10. & 27 & 7 & 6 & 64 & 350 & 500-2800 & 4-12 \\ \hline
100 & 3.0 & 10.6 & 13 & 7 & 6 & 64 & 330 & 500-2800 & 4-12 \\
143 & 2.1 & 7.4 & 12 & 6 & 4 & 56 & 290 & 330-3600 & 5-15 \\
217 & 1.38 & 4.9 & 14 & 5 & 4 & 31 & 180 & 350-4000 & 120-200 \\
353 & 0.85 & 4.5 & 24 & 18 & 18 & 16 & 200 & 250-2000 & 2000-3500 \\
545 & 0.55 & 4.5 & 44 & 62 & 45 & 3 & 450 & 40-250 & 6000-10000 \\
857 & 0.35 & 4.5 & 36 & 120 & 70 & - & 700 & 10-50 & 17000-25000 \\ 
& & & & & & & & & \\ \hline
\end{tabular}
\begin{description}
\item[${}^{(1)}$] Flux density fluctuations of the cosmic microwave 
background corresponding to $\Delta T/T = 10^{-5}$
\item[${}^{(2)}$] Adopted detection limit for discrete sources, 
equal to $5\times (\sigma^2_{\rm noise} + \sigma^2_{\rm conf} +  
\sigma^2_{\rm Galaxy}+\sigma^2_{\rm CMB})^{1/2}$. 
A rough estimate of anisotropies, $\sigma_{\rm Galaxy}$, due to
the galactic emission at high galactic latitudes ($\vert b\vert >20^\circ$),
has been obtained following Danese et al. (1996) and
Bersanelli et al. (1996; Fig. 2.3, 2.4 and Appendix A.1)
and taking into account
the angular dependence of galactic dust fluctuations determined by
Gautier et al. (1992). If we neglect this correction factor,
the galactic dust fluctuations at 217, 353, 545 and 857 GHz turn out
to be larger by a factor $\sim 2.2$.
\end{description} 
\end{table*}


\subsection{Total counts}
Our estimates (see Table 2 and Figure 4) indicate that Planck counts 
are dominated by radiosources for frequencies up to about 200 GHz. 
Note that the large differences in the estimated numbers of radio
selected sources is determined by the variation of the adopted values for
the average spectral index of compact sources (see Section 3.1).
It is interesting that significant numbers of radio galaxies
should be detectable in all frequency channels, so that the Planck mission 
will allow to explore their poorly known high frequency spectra. 

Also, the number 
of galaxies that should be detected in the high frequency channels 
is large enough to allow statistical investigations of their properties 
in this particularly interesting range and the definition of 
reliable local luminosity functions.

\section{Results and discussion}

Our estimates of the rms temperature fluctuations 
($(\Delta T/T)_{\rm rms}\equiv\langle (\Delta T/T)^2\rangle^{1/2}$) 
as a function of the angular scale,  
and of the angular power spectra [$\delta T_{\ell}$, eq.~(\ref{eq:dT})] 
of fluctuations 
due to extragalactic point sources are shown in Figures 5--6 and 7--8
respectively.

\subsection{Poisson fluctuations}

In Figures 5 and 6 the thick solid line corresponds to the case of 
``flat''-spectrum radio sources 
with average spectral index $\alpha =0$ above 20 GHz 
and far-IR sources evolving 
according to the updated model C by Franceschini et al. (1994). 
Sources brighter than 1 Jy are assumed to be individually identified and 
removed (a rather conservative assumption: cf. Table 2). 
The effect of subtracting fainter sources, identified by means 
of independent, deeper surveys is also shown. 

Although we expect that 
the individually detected sources in the high galactic latitude Planck 
surveys at frequencies up to about 200 GHz (see Table 2) will be mostly 
radio sources, a significant contribution to fluctuations may come 
from far-IR sources well below the detection limit.

In fact, the steepness 
of the counts of far-IR sources in the flux density interval where 
evolution sets in (and its effect is boosted by the steep K-correction) 
implies that their dominant contribution to fluctuations may not come, 
as usual, from flux densities corresponding to one source per beam (where 
radiosources probably dominate) but, in some cases, 
from the fainter fluxes where the slope $\beta$ of differential 
counts becomes $> 3$.

On an angular scale $\theta_0 \simeq 10'$ far--IR sources do not contribute
more than a few \% of the confusion fluctuations for 
$\nu \lsim 100\,$GHz; at 150--200 GHz the contributions of radio and far-IR 
sources are comparable, while at higher frequencies, far-IR sources dominate.

The amplitude of confusion fluctuations, after subtraction of only those 
sources which are individually detected by the Planck instruments themselves, 
decreases from $\Delta T/T\simeq 5\times 10^{-6}$ at 30 GHz to 
$\simeq 10^{-6}$ at $\sim 100$--150$\,$GHz, well below 
the expected rms primordial fluctuations.
 
It may be noted that up to $\simeq 100\,$GHz the rms fluctuations due to 
discrete sources, as a function of the angular scale, show a broad maximum 
at $\theta_0 \simeq 10'$, while they increase with decreasing $\theta_0$ 
at higher frequencies, due to the contribution of evolving far-IR sources 
which becomes more and more important with increasing angular resolution of the 
survey. 

As previously pointed out, the estimates of fluctuations due to far-IR 
sources are particularly uncertain since the available information 
on both their spectra and their evolutionary properties is poor. 
However, the deep IRAS $60\,\mu$m counts on one side and 
the upper limits on the far-IR background intensity on the other, 
strongly constrain the amplitude of fluctuations at the angular resolution 
of the high frequency channels of the Planck mission.
Phenomenological pure luminosity evolution models without changes
in the source spectra with cosmic time (Franceschini et al., 1988; 1991)
consistent with the high $60\,\mu$m counts by Gregorich et al. (1995), 
which may be overestimated at the faint end because of source confusion, 
and barely consistent with the conservative upper limit on the extragalactic 
far-IR background derived by Shafer et al. (1997), entail fluctuations 
very close to those implied by the updated model C on scales above the 
angular resolution of the Planck mission; only below $\sim$1.5 arcmin the
fluctuation level increases by a factor of 1.5--2.

Note that the effect of source variability is already included, in a
statistical sense, in the above estimates. In fact, variability affects
source counts in a manner similar to the Eddington effect: the observed
counts are made sistematically higher at brighter fluxes, since sources
are preferentially detected in their brighter phases. Since fluctuations
are estimated from observed counts, the effect is automatically taken
into account. On the other hand, variability seriously hinders subtractions
of sources based on observations at different frequencies or 
non--simultaneous.

\subsection{The effect of clustering} 

In the case of radio sources, based on the results by Loan et al. 
(1997), we have adopted an angular two-point correlation function 
$w(\theta)=0.01$ for $\theta = 1^\circ$. For the double power law 
representation of $\xi (r,z)$, described in Section 2.2, this implies a 
clustering radius $r_0 = 36 (50/H_0)\,$Mpc if $\epsilon = 0$ (stable 
clustering); if $\epsilon = -1.2$ (clustering constant in comoving 
coordinates), $r_0 = 26 (50/H_0)\,$Mpc. The two models yield differences 
on the estimated contribution of clustering to 
fluctuations of 10--15\%. The results shown in Figs. 5 and 6 refer to 
$\epsilon = -1.2$.

For far--IR sources we have adopted $r_0=10\,$Mpc ($H_0=50$)  
and $\epsilon=0$. This value of the clustering radius is slightly 
larger than that estimated by Saunders et al. (1992) 
for far--IR selected sources, on account of the fact that, according to 
our reference model, a substantial contribution to fluctuations comes 
from early phases of the evolution of early type galaxies, which are 
more clustered than the late type galaxies studied by Saunders et al. (1992). 

The adopted evolutionary and spectral properties of the sources are the same 
as for the calculations of Poisson fluctuations. In all cases the 
latter dominate except in the case 
that sources can be identified and subtracted down to fluxes below 
the detection limit of the survey, thus greatly decreasing the Poisson 
fluctuations, while the contribution due to clustering is only weakly 
affected.

As shown by Fig. 6, the autocorrelation of intensity 
fluctuations essentially 
reflects the clustering of radiosources below 200 GHz and 
of far-IR sources at higher frequencies; around 200 GHz the contributions 
of the two populations are similar. 

\subsection{Angular power spectra of fluctuations}

Following Tegmark \& Efstathiou (1996) we have computed, and plotted in
Figures 7 and 8, the quantity $\delta T_{\ell}(\nu) = 
\left[\ell(\ell + 1)C_{\ell}(\nu)/2\pi\right]^{1/2}$ 
which is roughly the average rms temperature fluctuation in the 
multipole range $\ell_0\leq \ell \leq \ell_1$ with $\ln(\ell_1/\ell_0)=1$.

The plotted power spectra are the mean of 50 simulations of the all sky
distribution of sources. Fluctuations are dominated by radio sources 
at the lowest frequencies (30, 44, 65 and 100 GHz) and by the far-IR sources 
at the highest (353, 545 and 857 GHz); both classes of sources 
are important at the intermediate frequencies (143, and 217 GHz).
For far-IR sources the updated model C has been adopted. As mentioned above, 
this model may overestimate the fluctuation level. A spectral index 
$\alpha=0$ was adopted for ``flat-spectrum'' radiosources. The effect of 
clustering was neglected.

It is assumed that sources brighter than 1, 0.1 and 0.01 Jy are identified
and removed from the maps (see caption). For the 217 GHz band we also show the 
effect of removing only the few brightest sources ($S> 10\,$Jy).  
As shown by Table 2, we
expect that it will be possible to directly detect, and subtract out, 
sources down to flux densities of a few to several hundred mJy.

It may be noted that lowering the flux limit for source subtraction 
decreases the amplitude of fluctuations more effectively at low than at 
high frequencies. This is because of the steepness of the counts 
of far-IR sources at sub-mm wavelengths, implying substantial contributions 
to fluctuations from very faint sources.

Our results for a flux density cutoff of 100 mJy
are substantially below those obtained by Tegmark \& Efstathiou (1996)
adopting the same flux limit (see Fig. 7). This is due to
their assumption of a spectral index $\alpha = 0$ for {\it all sources} 
selected at 1.5 GHz. This leads to a substantial overestimate of the 
expected counts, since most 1.5 GHz sources in the 
relevant flux density range are known to have a steep ($\alpha \simeq 0.7$) 
spectrum. 

\section{Conclusions}

Although, as stressed above, the present estimates are rather uncertain, 
particularly in the high frequency bands, it can be safely concluded that 
at least the central frequency channels will 
allow a clean view of primordial anisotropies up to the 
maximum $\ell$--values accessible to the Planck mission. In fact, even 
under the most extreme assumptions compatible with the available information 
on high frequency spectra of radio sources and on the evolution properties 
of far-IR sources, the amplitude of fluctuations due to discrete sources in 
the 100--200 GHz range are well below the expected amplitude of primordial 
anisotropies. 

The availability of multifrequency data allows an efficient identification of 
pixels contaminated by discrete sources.
Table 2 shows that many sources not directly detectable in the 
central frequency channels can be identified at higher or lower frequencies. 
The corresponding pixels can be simply removed; thanks to the large area 
surveyed, the number of remaining clean pixels will be nevertheless very high. 
But a reliable subtraction 
of the contaminating flux may also be possible thanks to the fact that a 
significant number of sources of the various 
classes should be detectable in most or all frequency channels, 
allowing the definition of well defined template spectra. 

Moreover, removal of contaminating signals is eased by the substantial 
difference between their power spectrum and that of primordial fluctuations 
(Tegmark \& Efstathiou 1996).

On the other hand, the Planck mission will bridge 
the gap between radio surveys, carried out at 
$\nu \leq 8.4\,$GHz ($\lambda \geq 3.6\,$cm) and far-IR surveys (IRAS and ISO) 
at $\nu \geq 1500$--3000$\,$GHz ($\lambda \leq 100$--$200\,\mu$m), 
providing the first exploration of the whole sky 
in a spectral region where many interesting 
astrophysical phenomena are most easily investigated. 
A pot-pourri 
of issues for which these data will be extremely relevant include: 
bremsstrahlung emission as a tracer of evolution of stellar populations; 
high-frequency behaviour of the spectra of compact radio sources and 
implications for their physical properties; definition of unbiased 
samples of blazars; cold dust in galaxies and hints on its evolution; 
physical and evolutionary connections between nuclear 
activity and processes governing the abundance and the properties 
of the interstellar material; the relationships between different AGN 
classes and tests for unified models; 
energy source(s) of the huge far-IR emission from 
type 2 Seyferts and from some QSOs and radiogalaxies. 

\section{ACKNOWLEDGMENTS}
We wish to thank an anonymous referee for his comments and suggestions
which helped us to improve the final presentation of this paper. 
We are grateful to M. Bersanelli and N. Mandolesi for providing us
with updated information on the Low and High Frequency Instruments
(LFI and HFI) foreseen for the Planck mission and to
G. Smoot for useful comments on an early draft of this paper.
This work has been partially supported by the Consiglio Nazionale
delle Ricerche (CNR) and by the Agenzia Spaziale Italiana (ASI), contract
95--RS--116. LT and FAG would like to thank
the Vicerrectorado de Investigaci\'on of the University of Oviedo (Spain)
for continuous financial support during the years 1995 and 1996 (projects
DF/95--213--1 and DF/94--213--6). LT and FAG acknowledge partial
financial support from the Spanish ``Direcci\'on General de Ense\~nanza
Superior'' (DGES), under project PB95--1132--C02--02.

\newpage
\medskip
\centerline{\bf FIGURE CAPTIONS}

\medskip\medskip
{\bf Figure 1.} Comparison between predicted and observed differential source 
counts at 1.4, 5 and 8.44 GHz normalized to
$150 S^{-2.5}\,\hbox{sr}^{-1}\,\hbox{Jy}^{-1}$ (see top panel).
The contributions of the most relevant classes
of radio sources according to the model of Danese et al. (1987) are shown.
For references for the data points see Danese et al. (1987);
additional data are from Donnelly et al. 
(1987), Fomalont et al. (1991), Windhorst et al. (1993).

\medskip\medskip
{\bf Figure 2.} Differential $60\,\mu$m galaxy counts normalized to 
$600 S^{-2.5}\,\hbox{sr}^{-1}\,\hbox{Jy}^{-1}$. Data are from Hacking \& 
Houck (1987), Rowan-Robinson et al. (1991), Gregorich et al. (1995), and 
Bertin et al. (1997). The solid line shows the predictions of 
model C by Franceschini et al. (1994), updated as described in the text. 
Also shown are the contributions to the counts from late type galaxies 
(spiral+irregular, short dashed line; starburst, dotted line), from
early type galaxies (E and S0), assumed to go through a strongly dust
absorbed phase during their early evolution (dot-dashed line),
and from AGNs (long/short dashes).

\medskip\medskip
{\bf Figure 3.} Integral counts at $170\,\mu$m and at $850\,\mu$m. 
The meaning of the lines is the same as in Figure 2. The data points 
are the preliminary estimates by Kawara et al. (1997) at $170\,\mu$m 
and by Smail et al. (1997) at $850\,\mu$m. As stressed by the latter 
authors, their result might be, conservatively, taken as an upper limit; 
the plotted error bars for this point are $2\sigma$.

\medskip\medskip
{\bf Figure 4.} Predicted integral counts in the Planck bands. The solid 
line shows the total counts. The dashed and dotted lines show, respectively, 
the contributions of radio and far--IR selected sources, as 
predicted by the models 
by Danese et al. (1987) and the Franceschini et al. (1994), updated 
as described in the text. We have adopted an average spectral index 
$\alpha =0$ for ``flat''-spectrum radio sources and the ``opaque'' model 
for the early evolution of early-type galaxies in the far-IR.

\medskip\medskip
{\bf Figure 5.} Temperature fluctuations 
due to discrete sources at $\nu$=100 GHz as a function of the angular 
scale. The solid, short-dashed, long-dashed, and dot-dashed lines 
covering the full range of angular scales correspond to the estimated 
Poisson contributions for different choices 
of the limiting flux above which sources are individually detected 
and subtracted out: 1 Jy, 0.1 Jy, 10 mJy, and 1 mJy, respectively. 
The lines plotted for scales larger than the Planck resolution at this 
frequency ($10'$) show the contributions due to clustering of radiosources 
(clustering of far-IR sources has a negligible effect at this frequency); 
the upper curve assumes that sources are subtracted for $S >1$ Jy, the lower 
one, for $S >0.1$ Jy. 
 
\medskip\medskip
{\bf Figure 6.} Temperature fluctuations due to discrete extragalactic 
sources at the frequencies indicated in each panel, as a function of the 
angular scale. As for the Poisson contribution,
the lines have the same meaning as in 
Figure 5. Again, the contributions due to clustering are shown only for 
scales larger than the angular resolution of Planck instruments at each 
frequency: the dotted and dot-dashed lines correspond to radiosources and
to a flux limit of 1 and 0.1 Jy for source subtraction, respectively;
the solid line corresponds to far-IR sources and a to S$_{l}$=1 Jy
for source subtraction (lowering S$_{l}$ down to 0.1 Jy provides
a negligible variation of the estimated contribution).
At frequencies up to $\sim$ 100 GHz the effect of clustering of far-IR
sources is negligible and only the contribution of radiosources is shown; 
the opposite is true at $\nu \geq 353\,$GHz. The contributions of 
both populations are given at 143 and 217 GHz; at the latter frequency 
they are very close to each other. 

\medskip\medskip
{\bf Figure 7.} Angular power spectra of fluctuations due to a Poisson 
distribution of discrete sources at $\nu=100\,$GHz (see text for details).  
The solid, dot-dashed, and long/short dashed lines correspond to a flux 
limit for source removal of 1, 0.1, and 0.01 Jy, respectively. 
The dotted lines show the ``unsmoothed'' noise fields 
(defined as in Tegmark \& Efstathiou 1996) foreseen for 
the Planck Surveyor detectors. 
The roughly horizontal thin dot--dashed line shows the dependence on multipole 
of temperature fluctuations predicted by the standard CDM model 
(scale-invariant scalar fluctuations in a $\Omega=1$ universe with $H_0=
50\,\hbox{km}\,\hbox{s}^{-1}\,\hbox{Mpc}^{-1}$, and baryon density 
$\Omega_b=0.05$).
The thin long--dashed line, labelled T\&E96, shows the power spectrum 
estimated by Tegmark \& Efstathiou (1996).

\medskip\medskip
{\bf Figure 8.} Same as in Figure 7, for the frequencies indicated in each 
panel. The dashed line in the 217 GHz panel refers to a removal of only the 
few brightest sources ($S > 10\,$Jy). 
The short--dashed line in the upper left corner of the 30 GHz panel  
shows, for comparison, the ``unsmoothed'' noise field 
for COBE/DMR, calculated adopting the average 
pixel noise given by Bennett et al. (1994). 

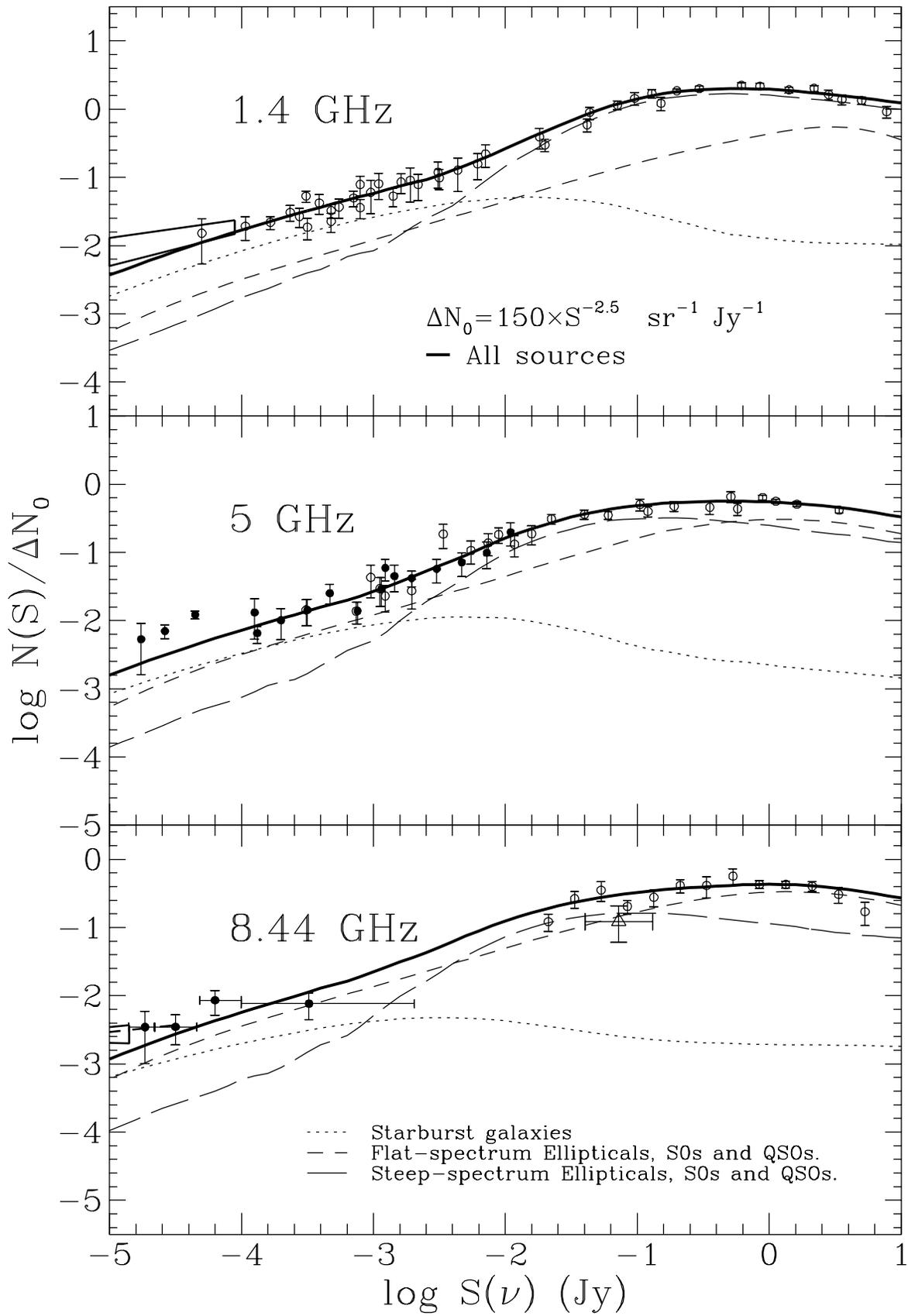
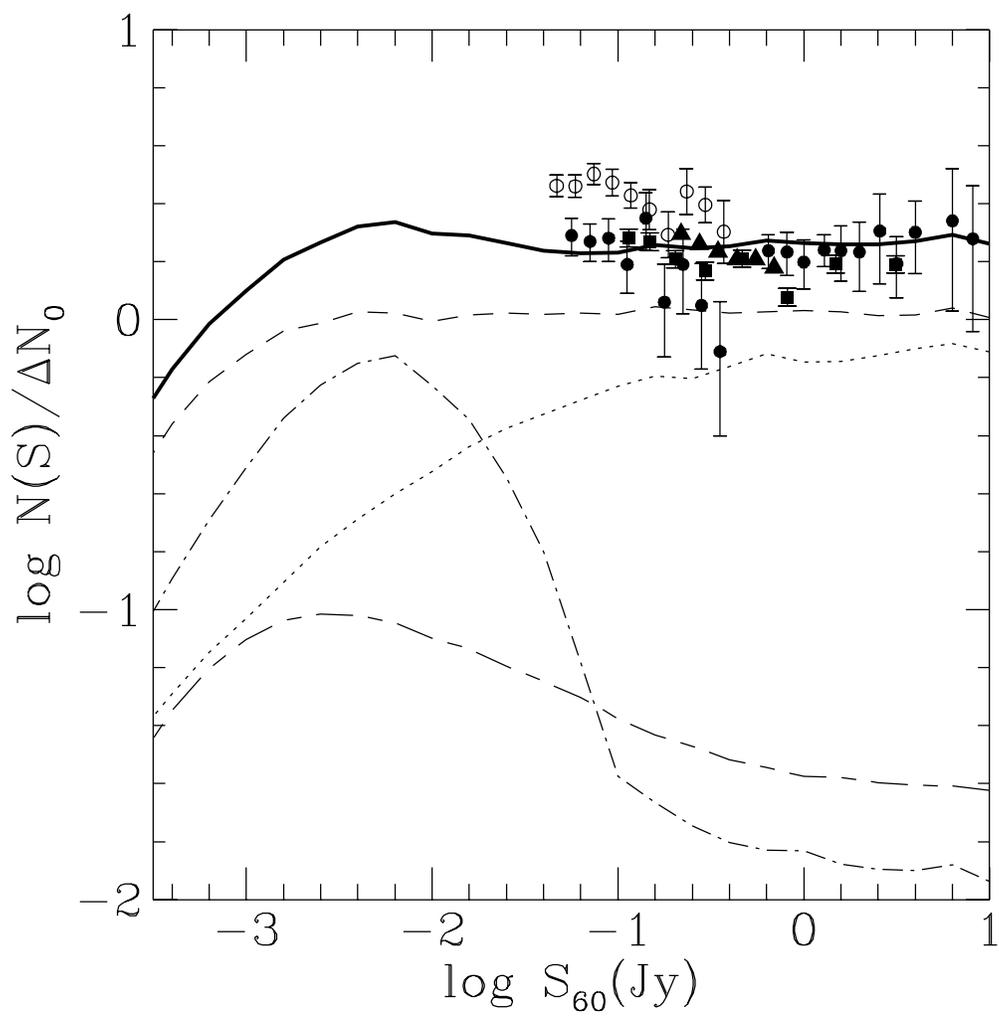
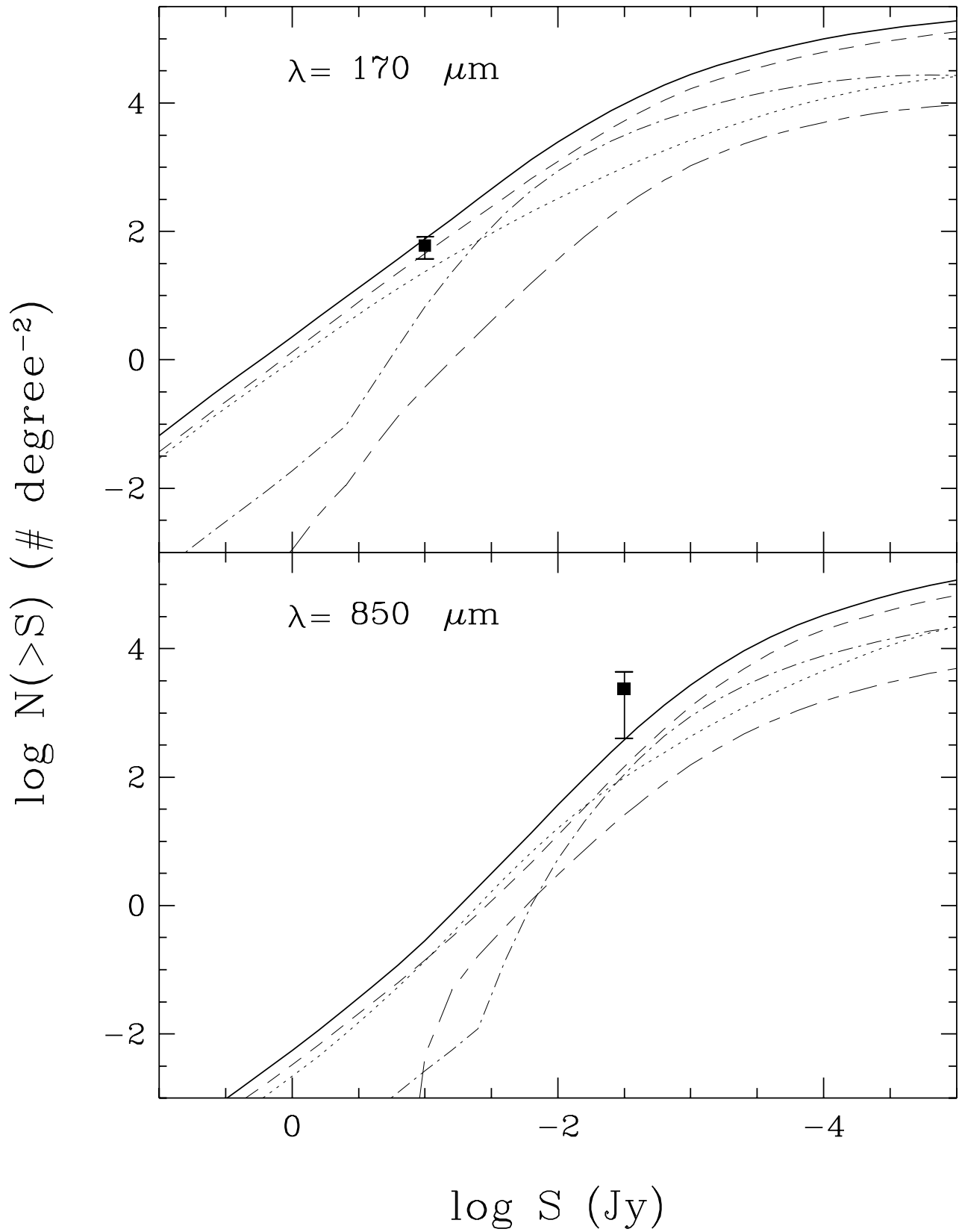
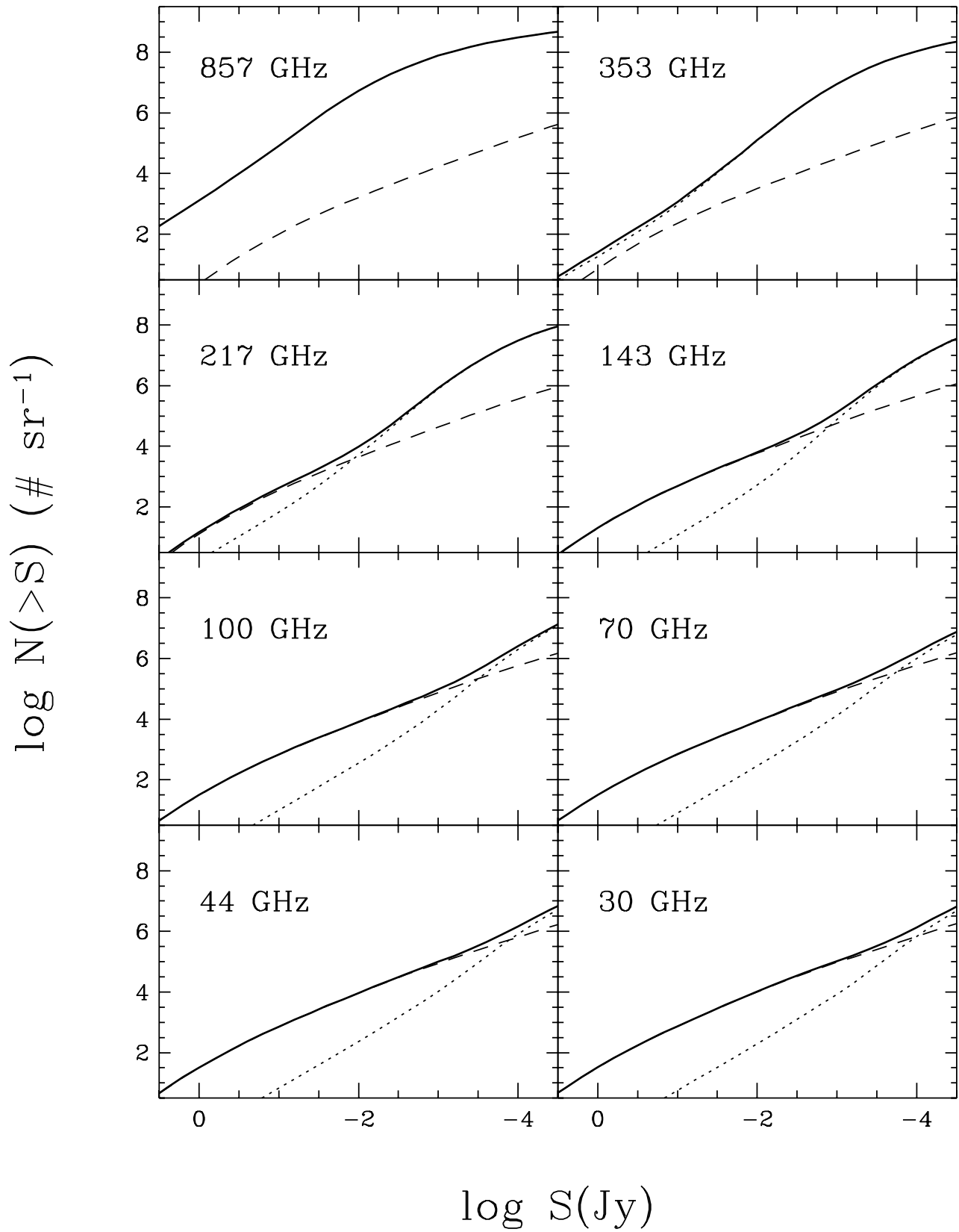
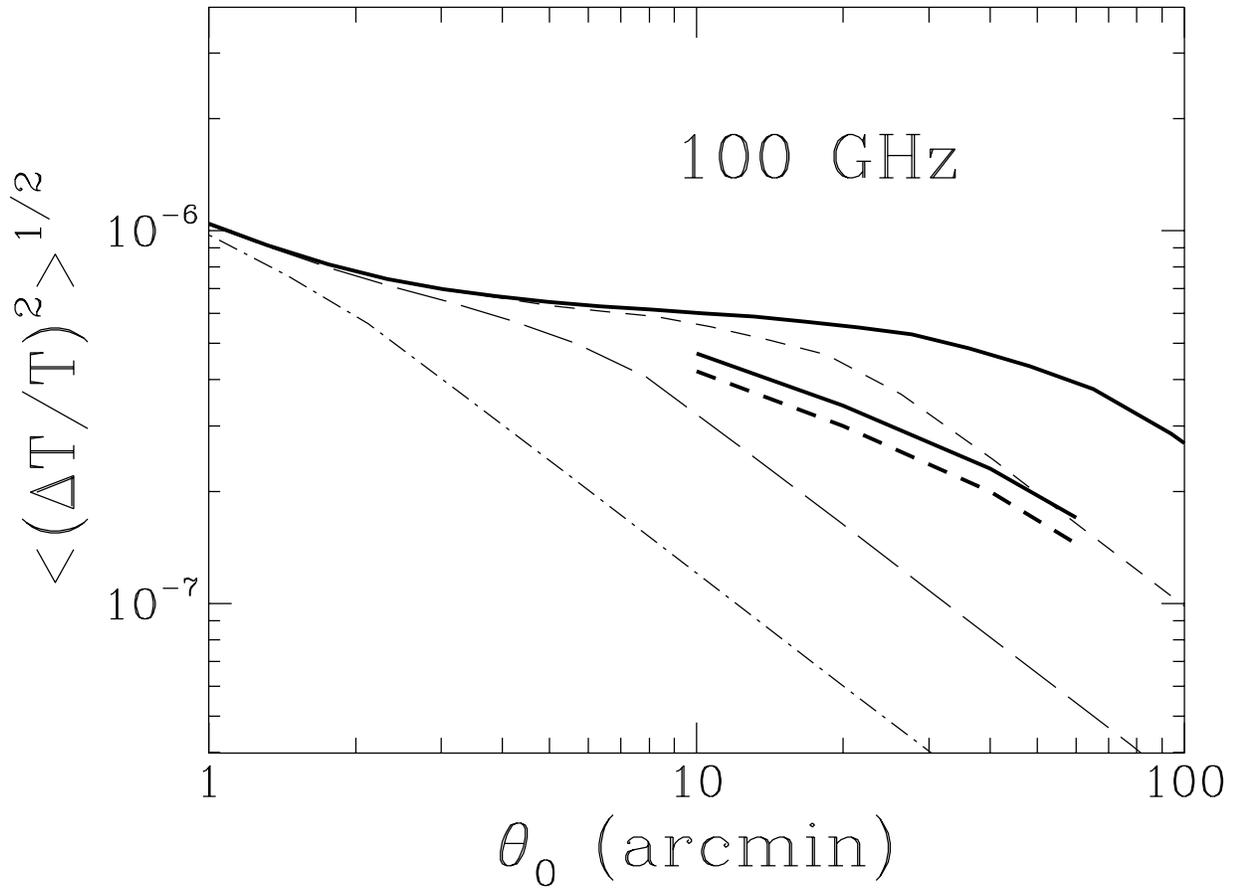
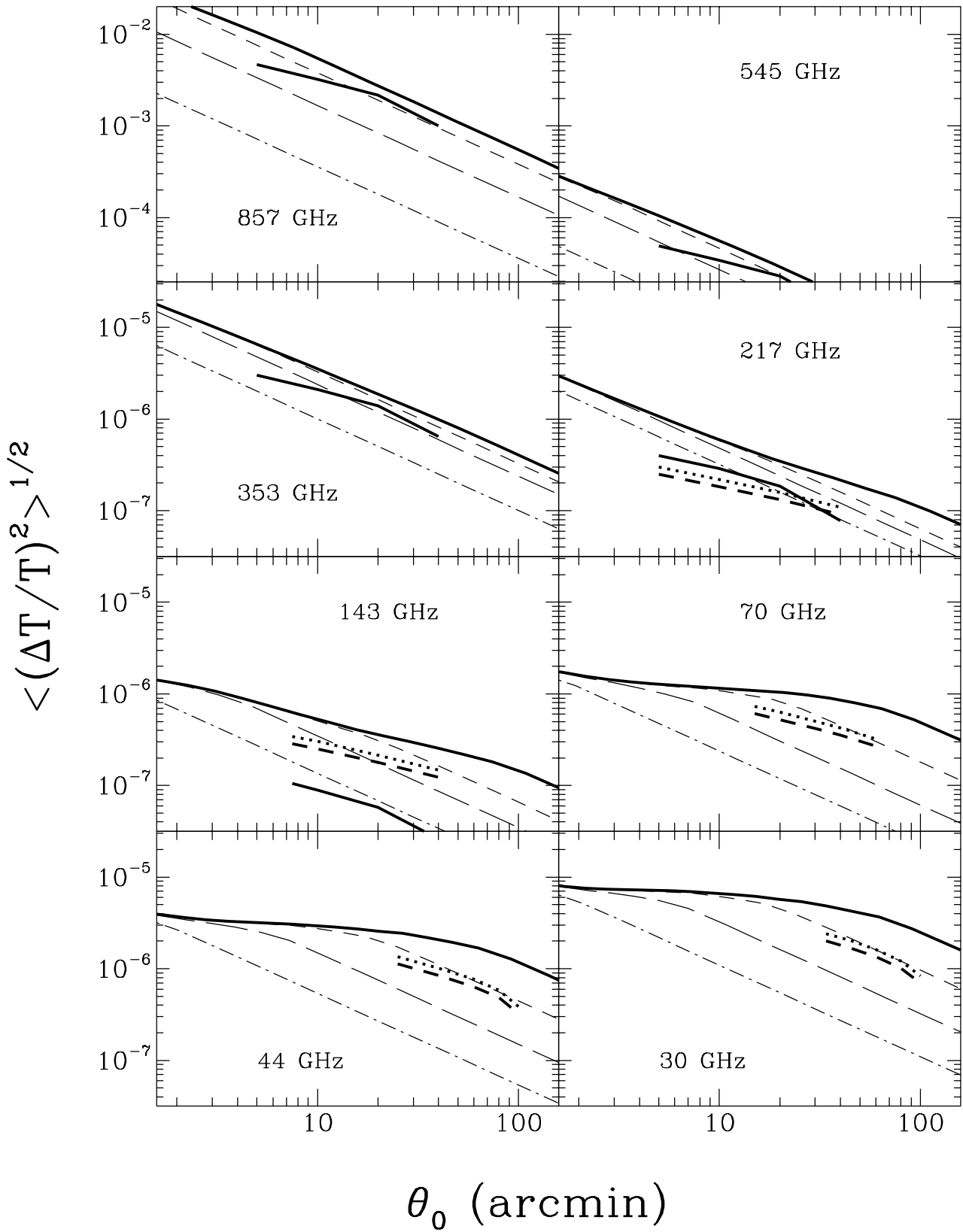
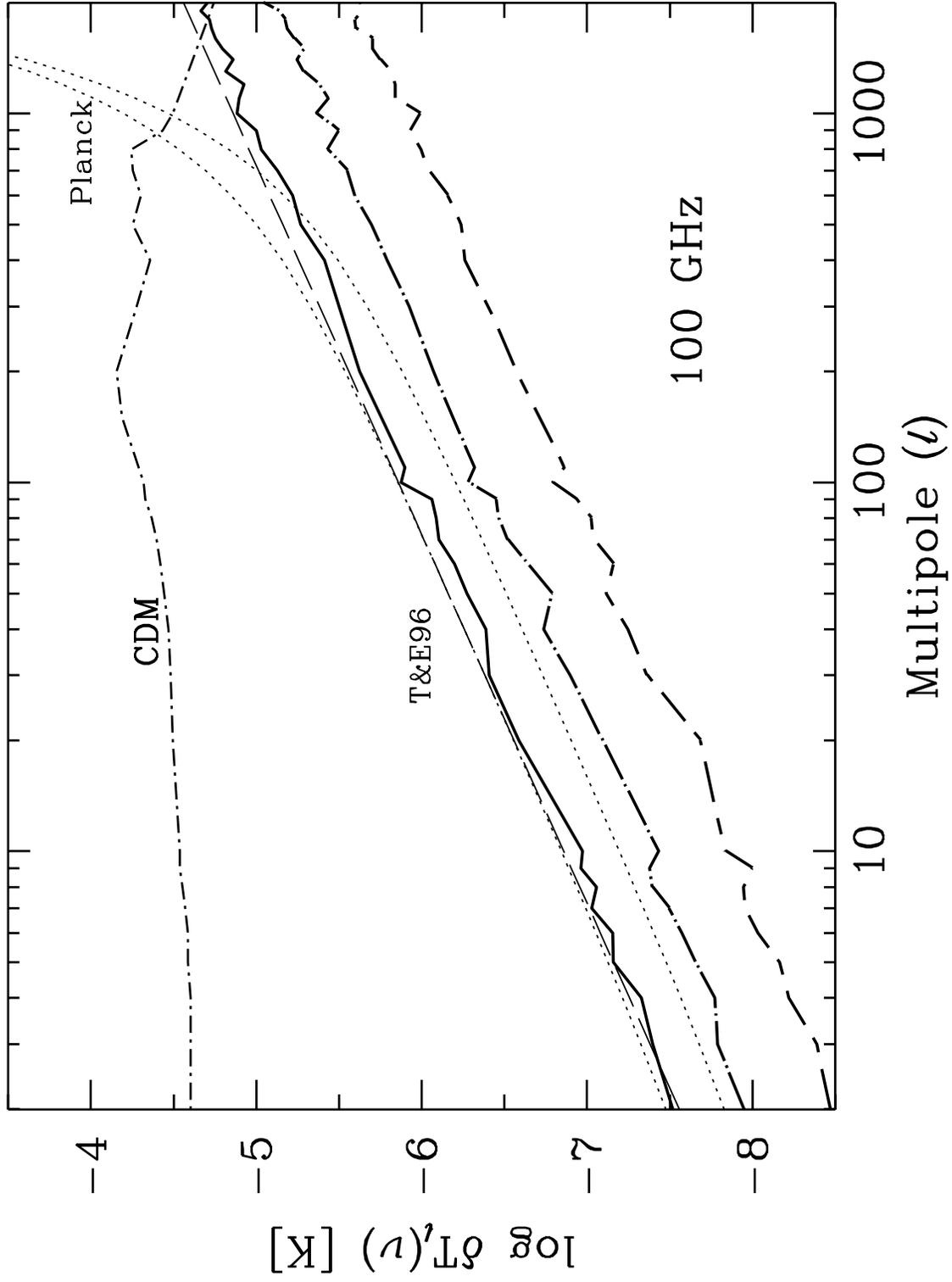
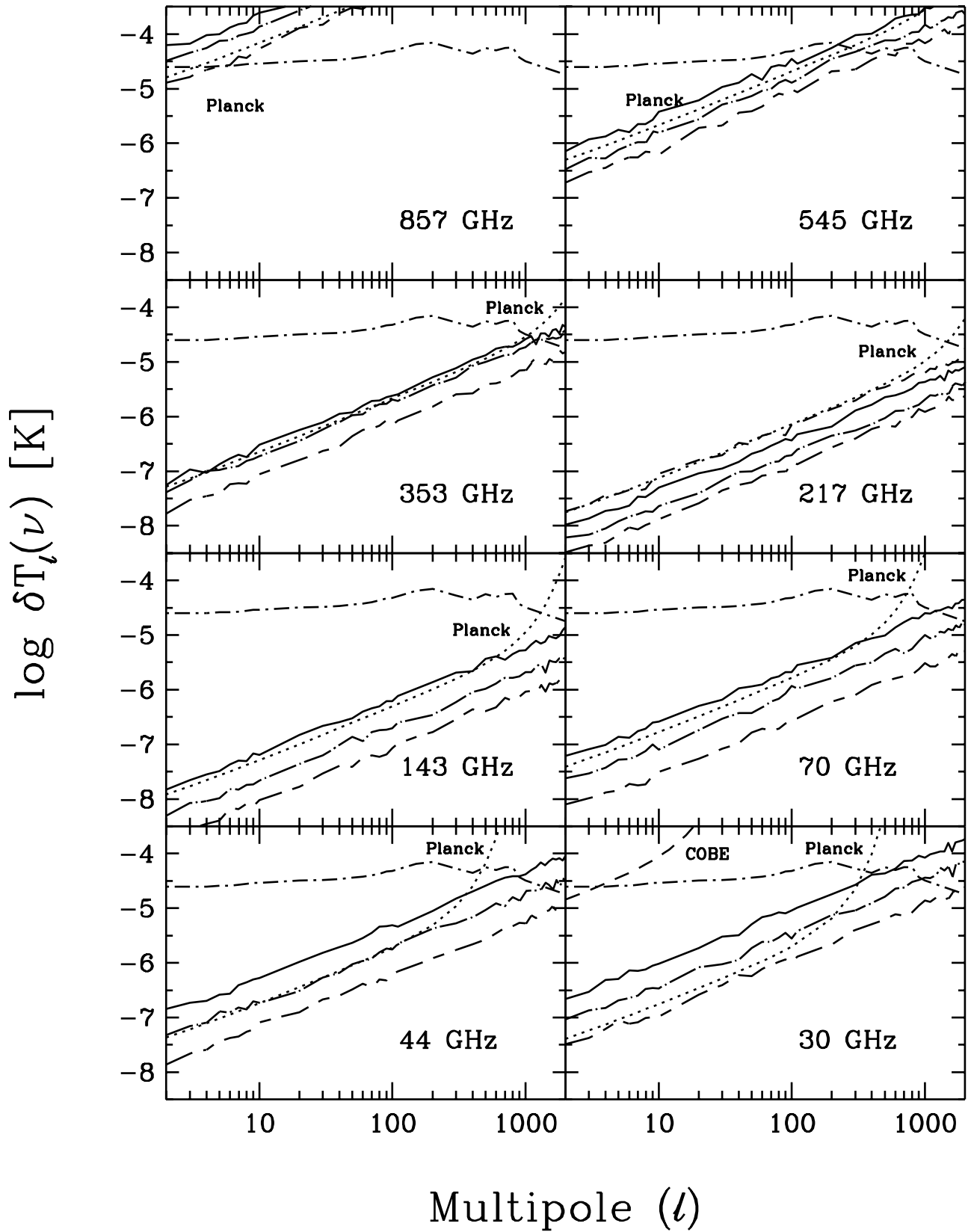
\end{document}